\documentclass[onecolumn,showpacs,preprintnumbers,amsmath,amssymb] {revtex4}
\usepackage{graphicx}% Include figure files
\usepackage{dcolumn}% Align table columns on decimal point
\usepackage{bm}% bold math

\begin{document}

\title{Spinons and triplons in spatially anisotropic frustrated
   antiferromagnets}

\author{Masanori Kohno$^{1,2}$, Oleg A. Starykh 
$^{3}$, and Leon Balents$^{1}$}
\affiliation{$^1$Department of Physics, University of California,  
Santa Barbara, CA 93106, USA\\
$^2$Computational Materials Science Center, National Institute for  
Materials Science, Tsukuba 305-0047, Japan\\
$^3$Department of Physics, University of Utah, Salt Lake City, UT  
84112, USA}

\date{\today}

\begin{abstract}
  The search for elementary excitations with fractional quantum
  numbers is a central challenge in modern condensed matter physics.
  We explore the possibility in a realistic model for several
  materials, the spin-1/2 spatially anisotropic frustrated Heisenberg
  antiferromagnet in two dimensions.  By restricting the Hilbert space
  to that expressed by exact eigenstates of the Heisenberg chain, we
  derive an effective Schr\"odinger equation valid in the weak
  interchain-coupling regime.  The dynamical spin correlations from
  this approach agree quantitatively with inelastic neutron
  measurements on the triangular antiferromagnet Cs$_2$CuCl$_4$.  The
  spectral features in such antiferromagnets can be attributed to two
  types of excitations: descendents of one-dimensional spinons of
  individual chains, and coherently propagating ``triplon'' bound
  states of spinon pairs.  We argue that triplons are generic features
  of spatially anisotropic frustrated antiferromagnets, and arise
  because the bound spinon pair lowers its kinetic energy by
  propagating between chains.
\end{abstract}


\maketitle

One of the most dramatic effects of strong interactions in electronic
materials is the emergence of particles with fractional quantum
numbers, for example charge $e/3$ Laughlin quasiparticles in the
fractional quantum Hall efect, and spin-charge separated excitations
in one dimensional (1D) quantum wires and carbon nanotubes.  Indeed,
fractionalization is known to be quite generic in 1D conductors and
magnets.\cite{TL1,TL2} In this case the spin excitation carrying a
fractional quantum number, spin $1/2$, is referred to as a spinon
\cite{FT,HSspinon}.  In contrast, in dimensions higher than one, the
elementary excitation from a magnetically ordered states is known as a
magnon and carries spin 1 \cite{magnon,SWAF1,SWAF2}. Nevertheless,
fractionalization caused by strong quantum fluctuations has been
repeatedly identified theoretically as a possible phenomena underlying
unusual experimental behavior of strongly correlated materials in two
and three dimensions and zero magnetic field, such as high-temperature
superconductors, heavy fermions, and frustrated quantum magnets.  In
these contexts, resonating valence bond (RVB) theories
\cite{RVB,RVBexcitation} and slave-particle approaches
\cite{SlaveBosonHubbard,SlaveBosonHighTc,HighTcRev} have been
developed to describe fractionalization in dimensions greater than
one\cite{deconfinement}.  However, these approaches remain largely
unproved.  Considerable effort has been devoted to the search for such exotic
behaviors for decades \cite{PIRG1,PIRG2}, and only recently,
experimental indications of fractionalized particles
\cite{ColdeaPRL,ColdeaPRB} and disordered ground states
\cite{BETT-TTF,He3,Kagome1,Kagome2} have been observed in highly
frustrated antiferromagnets in two dimensions (2D).

\begin{figure} \includegraphics[scale=0.18]{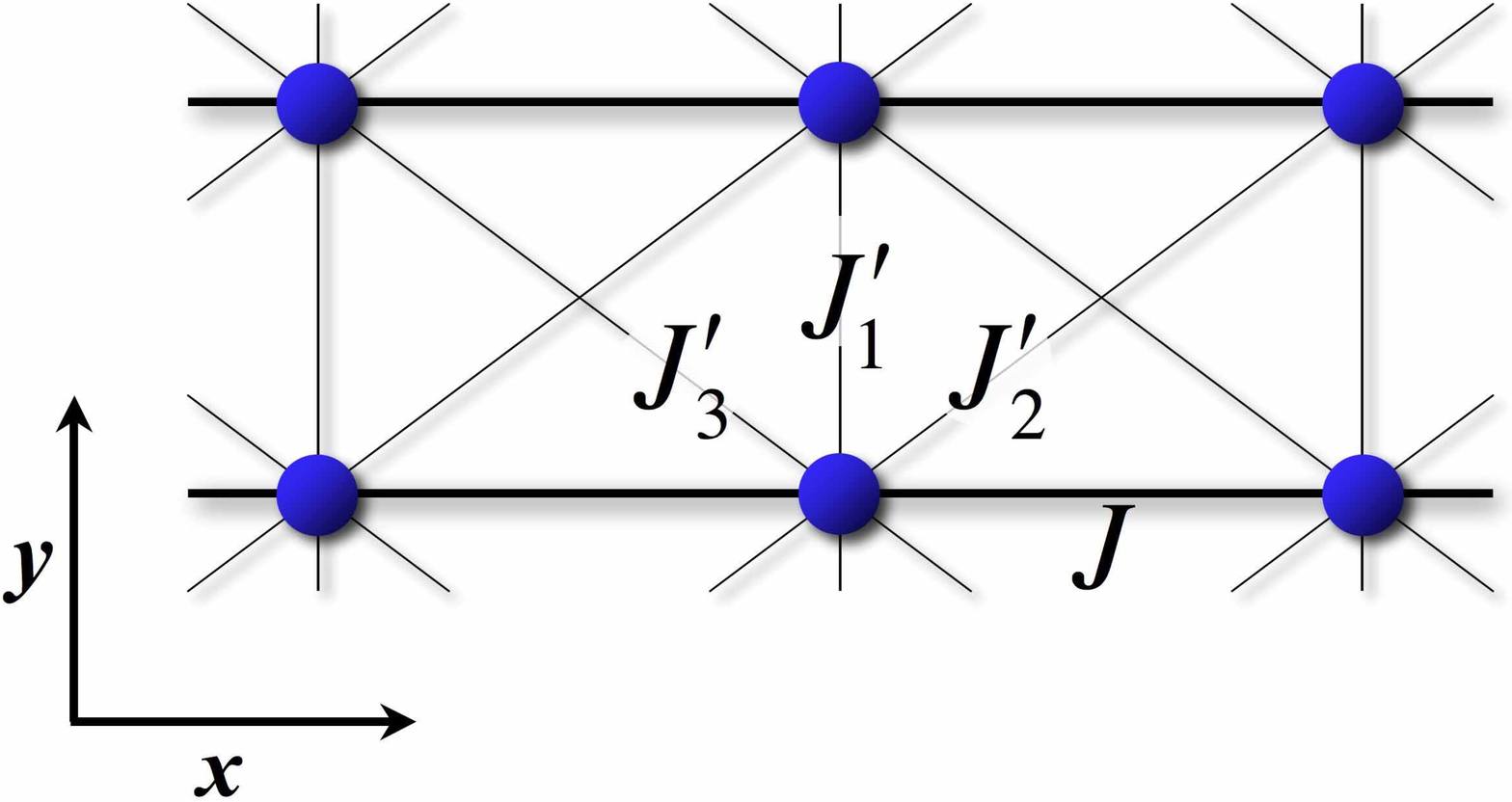} \caption{Lattice
structure and coupling constants $J'_1$, $J'_2$, $J'_3$ and $J$. Dots
and lines denote sites and bonds, respectively.}  \label{fig:lattice}
\end{figure}

In this paper, we consider how spinons may appear in a 2D
magnet as descendents of their 1D counterparts.  Our
focus is the spin-1/2 spatially anisotropic
antiferromagnetic Heisenberg model defined by the following
Hamiltonian:
\begin{equation}
{\cal H}=\sum_{x,y}\left(J{\mbox {\boldmath $S$}}_{x+1,y}+J'_1{\mbox
     {\boldmath $S$}}_{x,y+1}+J'_2{\mbox {\boldmath
       $S$}}_{x+1,y+1}+J'_3{\mbox {\boldmath
       $S$}}_{x-1,y+1}\right)\cdot{\mbox {\boldmath $S$}}_{x,y},
\label{eq:1}
\end{equation}
where ${\mbox {\boldmath $S$}}_{x,y}$ is the spin-1/2 operator at site
$(x,y)$. Here, $J$ denotes the intrachain coupling, and $J'_1$, $J'_2$
and $J'_3$ are interchain couplings as illustrated in
Fig.~\ref{fig:lattice}. We take all the coupling constants positive,
reflecting antiferromagnetic interactions, focusing on the frustrated
situation $J'_1=J'_2+J'_3$.  The main result of this paper is a
systematic method to calculate the {\sl inelastic} magnetic structure
factor $S(k,\omega)$ {\sl for the full range of energy transfers} with
$\omega$ varying from essentially zero to large scales of several
times $J$.  The result is valid provided only $J'_a/J$ is not too large,
and indeed reveals characteristic features of spinon excitations.

One strong motivation to study this model comes from experiments on
the material Cs$_2$CuCl$_4$, a spin-$1/2$ Heisenberg antiferromagnet
on a spatially anisotropic triangular lattice.  This corresponds to
Eq.\eqref{eq:1} with $J'_1=J'_2\equiv J'$ and $J'_3=0$, and the {\sl
  measured} anisotropy is $J/J' \approx 3$.\cite{ColdeaModel} The
spectral weight in the dynamical structure factor, $S(k,\omega)$,
measured in this compound is dominated by a broad continuum, extending
up to energy above $3 J$, with the usually strong magnon peak
appearing uncharacteristically insignificant.  The spectral tail for
some directions in momentum space is well-fitted by a power-law
form\cite{ColdeaPRL,ColdeaPRB}. Following this observation, numerous
theories have attributed the behavior to fractionalized excitations of
exotic {\sl two dimensional} critical and/or spin liquid
states.\cite{wen02,AVL,O4SL} Other works have compared the data to
anharmonic spin wave theory.  Though the latter calculations reproduce 
the {\sl shape} of the observed dispersion of the
(broad) peaks in Cs$_2$CuCl$_4$, a substantial phenomenological
renormalization of the exchange parameters must be included {\sl by
  hand} to achieve quantitative
agreement.\cite{Spinwave1,Spinwave2,ColdeaPRL,ColdeaPRB} However,
numerical series expansion calculations using the un-renormalized
measured $J,J'$ values properly reproduce the experimental peak
dispersion.\cite{SeriesExp,Series07}

In this paper, we argue that the spectra in Cs$_2$CuCl$_4$ indeed
reflect the presence of spinon excitations as originally suggested,
but that these spinons are descendents of the 1D excitations of the
chains formed by the strong $J$ bonds, and not characteristic of any
exotic 2D state.  A popular argument {\sl against} this notion has
been that the peak energy has substantial dispersion in the direction
{\sl transverse} to the chains.  We show that contrary to na\"ive
expectations, such dispersion does appear in a quasi-1D approach.  The
basic physics involved is the {\sl binding} of two spinons into a
delocalized and dispersing spin-1 pair (triplon).  This is driven by
kinetic energy, since only a pair of spinons may hop between chains.
The idea is a lower dimensional analogue of Anderson's interlayer
tunneling mechanism of high temperature superconductivity, with spinon
pairs replacing Cooper pairs\cite{anderson88,highTc_intelayer}.
Triplon formation leads to specific signatures in the structure factor
{\sl which are indeed present in the data on Cs$_2$CuCl$_4$}.

 The appropriateness of the 1D approach is reinforced by several
 works.  Ref.\cite{triangDimer} demonstrated that it quantitatively
 reproduces most of the complex low temperature phase diagram observed
 in applied magnetic fields in Cs$_2$CuCl$_4$.  It also showed that
 the frustrated $J'$ coupling is ineffective in establishing
 long-range order: the characteristic energy scale for ordering is
 only of order $(J')^4/J^3$, much smaller than the bare $J'$
 inter-chain exchange energy.  An early indication of this
 ineffectiveness appeared in Ref.~\cite{singh99}, in which a
 ``decoupled'' state was suggested.  More recently, the exact
 diagonalization study in Ref.~\cite{sheng06} found that correlations
 between spins in neighboring chains remain extremely weak for $J'
 \leq 0.7 J$.

 This suggests that the elementary excitations (spinons) of
 independent spin chains are a natural basis.  We therefore {\sl
   project} the Hamiltonian in Eq.\eqref{eq:1} into the subspace of
 eigenstates of the 1D decoupled chains \cite{BetheAnsatz,1DHeisE}.  
 Each eigenstate can be
 characterized by the {\sl number of excited spinons}, which is always
 even for any physical state.  Remarkably, truncating to the first
 non-trivial approximation of only zero- or two-spinon states
 reproduces the main features of the spectrum of such
 quasi-one-dimensional frustrated antiferromagnets.  Note that the
 two-spinon approximation is {\sl not} a low-energy one (unlike the
 familiar and powerful ``bosonization'' technique) as it includes
 spinons with energies reaching up to $\pi J/2 \gg J'$.  This is
 essential for comparison with inelastic neutron scattering data which
 extends over this full range.\cite{ColdeaPRB}

The two-spinon states of a single chain are characterized by two
continuous quantum numbers, which can be thought of either as the
momenta $k_{x1},k_{x2}$ of the individual (unbound) spinons, or
equivalently, the total momentum $k_x=k_{x1}+k_{x2}$ and (excitation)  
energy
$\epsilon=\epsilon_s(k_{x1})+\epsilon_s(k_{x2})$ of the pair.
We use the latter notation for convenience.
The spinon energy is given by des Cloizeaux-Pearson dispersion,
$\epsilon_s(k_x) = (\pi J/2) |\sin(k_x)|$ \cite{dCP}.
   The states can also be
characterized by their total spin and $S^z$ quantum numbers.  Only the
triplet ($s=1$) states are relevant to the neutron structure factor, and
one may specialize without loss of generality to the  $S^z=+1$ state,
which we denote $|k_x,\epsilon\rangle_y$ on chain $y$.  For the
many-chain system, the unperturbed
ground state and two-spinon
basis states are given as $|{\rm
   G.S.}\rangle_0\equiv\otimes_y|0\rangle_y$ and
$|k_x,\epsilon,y\rangle\equiv |k_x,\epsilon\rangle_y\otimes_{y'\ne
   y}|0\rangle_{y'}$, respectively.  Here, $|0\rangle_y$ denotes the
ground state of the $y$-th Heisenberg chain, of length $L_x$.

We choose to work with eigenstates of the total 2D momentum vector ${\bf
   k}=(k_x,k_y)$.  Such $k_y$ eigenstates are  
superpositions: $|\epsilon\rangle_{\bf k}\equiv |k_x,k_y;
\epsilon\rangle\equiv \frac{1}{\sqrt{L_y}}\sum_y{\rm e}^{{\rm i} k_y y}
|k_x,\epsilon,y\rangle$ (here $L_y$ is the number of chains).  Note
that, because the two spinons comprising any of the original basis
states always live in the same chain, there is only one intrinsic
transverse momentum $k_y$ and not two distinct spinon momenta in the $y$
direction.  Thus there is only a one parameter ($\epsilon$) set of
two-spinon states for each $k_x,k_y$.  Therefore the eigenstates in this
basis take the form
\begin{equation}
|\Psi_{\bf k}\rangle =\int d\epsilon \, D_{k_x}(\epsilon)\psi_{\bf k} 
(\epsilon)
|\epsilon \rangle_{\bf k},
\label{eq:psi}
\end{equation}
where $D_{k_x}(\epsilon) = \Theta(\omega_{2,u}(k_x) - \epsilon) \Theta
(\epsilon - \omega_{2,l}(k_x))/ \sqrt{\omega_{2,u}^2(k_x) -
  \epsilon^2}$ is the density of states of the Heisenberg chain,
divided by $L_x/(2 \pi)$, at momentum $k_x$ and excitation energy
$\epsilon$ \cite{2spinonE} ($\Theta$ denotes the step function).  It
is restricted to $\omega_{2,l}(k_x) <\epsilon< \omega_{2,u}(k_x)$,
where the boundaries of the two-spinon continuum are $\omega_{2,l}(k)
= \epsilon_s(k_x)$ and $\omega_{2,u}(k_x) = \pi J \sin[k_x/2]$.  The
wavefunction $\psi_{\bf k}(\epsilon)$ defines the spread of the
eigenstate amongst this continuum.  The condition that $|\Psi_{\bf
  k}\rangle$ is an eigenstate of the Hamiltonian in the 2-spinon
subspace implies the Schr\"odinger equation:
\begin{equation}
   \label{eq:2}
   \epsilon \psi_{\bf k}(\epsilon) + \int\! d\tilde{\epsilon} D_{k_x} 
(\tilde\epsilon) \,  J'({\bf k}) A_{k_x}^*(\epsilon)A_{k_x}(\tilde 
\epsilon) \,\psi_{\bf
     k}(\tilde\epsilon)  = E\psi_{\bf k}(\epsilon),
\end{equation}
where $E$ is the excitation energy above the ground state, and $J'({\bf
   k})\equiv2(J'_1\cos k_y+J'_2\cos(k_x+k_y)+J'_3\cos(k_x-k_y))$ is the
Fourier transform of the interchain exchange interaction.  The matrix
element $A_{k_x}(\epsilon) \equiv
\frac{1}{\sqrt{2}}\langle0|S^-_{-k_x,y}|k_x,\epsilon\rangle_y$, which is
crucial for this study, was obtained exactly in Ref.~\cite{chainMqw}
(see Supplementary Material).

We solved the integral equation, Eq.\eqref{eq:2}, numerically by
carefully discretizing $\epsilon$ to obtain a complete (in the two
spinon subspace) set of eigenfunctions $\psi_{n{\bf k}}$ (and
corresponding states $|\Psi_{n{\bf k}}\rangle$) and energies $E_{n 
{\bf k}}$,
with $n=1\ldots M$.  Here we typically took the number
of discretized energies $M$ to be several thousand, as large as
necessary to ensure good resolution.  Knowing these eigenstates, we
can directly evaluate their contribution to the zero temperature
dynamical structure factor $S({\bf k},\omega)$:
\begin{equation}
\label{eq:9}
S({\bf k},\omega)=\int\frac{dt}{2\pi}{\rm e}^{{\rm i}\omega t}\langle
{\rm G.S.}|
S^\alpha_{-{\bf k}}(t)S^\alpha_{\bf k}(0)|{\rm G.S.}\rangle = \sum_n
\left|\langle {\rm G.S.}|S_{-{\bf k}}^\alpha
   |\Psi_{n{\bf k}}\rangle\right|^2 \delta(\omega-E_{n{\bf k}}).
\end{equation}
For consistency, we approximate the ground state $|{\rm G.S.}\rangle$
by its perturbative form to first order in $J'({\bf k})$, though the
linear correction term has little effect on the results.  Details are
given in the Supplementary Material.

Somewhat unexpectedly, it is possible to show analytically that the
structure factor obtained in this way has nearly the same form as
found in the well-known random phase approximation (RPA).  In
particular, as shown in the Supplementary Material, when the $O(J')$
correction to the ground state is neglected,
\begin{equation}
S({\bf k},\omega) = \frac{S_{1D}(k_x,\omega)}{[1 + J'({\bf k}) \chi_ 
{1D}'(k_x,\omega)]^2
+ [J'({\bf k}) \chi_{1D}''(k_x,\omega)]^2}.
\label{eq:RPA}
\end{equation}
Here $S_{1D}(k_x,\omega) = \chi_{1D}''(k_x,\omega)/\pi =
D_{k_x}(\omega) |A_{k_x}(\omega)|^2$ is the two-spinon structure
factor of a single chain \cite{chainMqwcalc}, and
$\chi_{1D}'(k_x,\omega)=\int_0^\infty \!  d\omega'
S_{1D}(k_x,\omega')/(\omega'-\omega)$.  This nearly coincides with the
RPA expression, which is obtained by replacing our $\chi$ with the
dynamic susceptibility of a single chain, $\chi_{1D}' \to {\rm
  Re}\chi_{1D}, \chi_{1D}'' \to {\rm Im}\chi_{1D}$.  ${\rm
  Re}\chi_{1D}$ differs from $\chi_{1D}'$ by a small contribution from
$\omega'<0$.  However, the differences between the RPA and our
two-spinon result, with or without the ground state correction, are
very small in all situations of interest -- see Supplementary
Material.

\begin{figure} \includegraphics[scale=1.0]{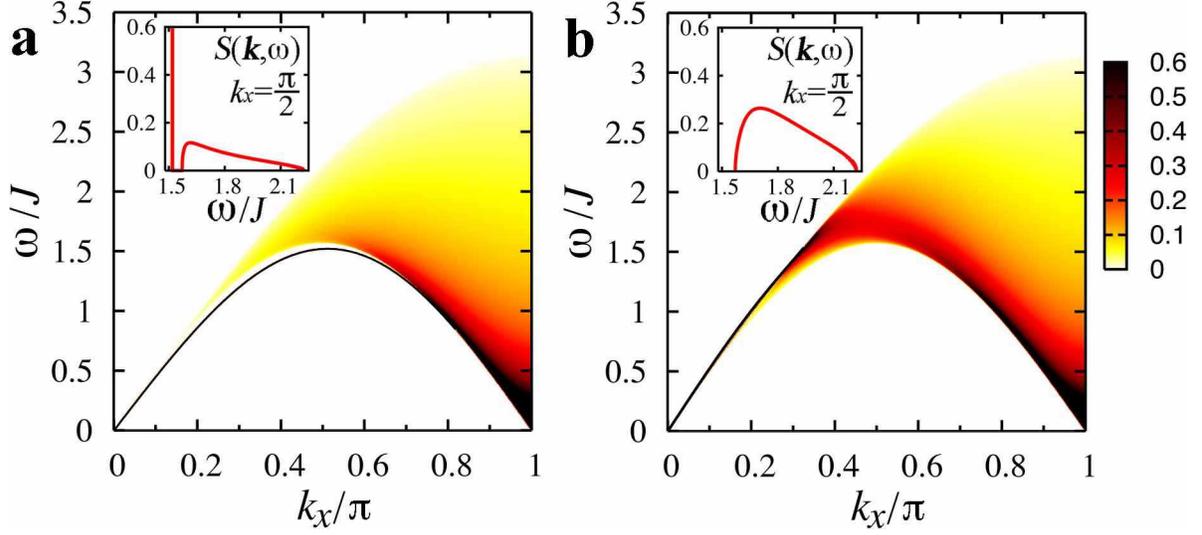} \caption{Density plot
of dynamical Structure factor $S({\bf k},\omega)$ for
$J'_2=J'_3=J'_1/2=0.24J$ at ({\bf a}) $k_y=\pi$ and ({\bf b})
$k_y=0$. The insets show the plots at $k_x=\pi/2$.}  \label{fig:skw}
\end{figure}

We find three types of distinctive spectral features depending on the
momentum, determined by the value of $J'({\bf k})$:
\begin{enumerate}
\item $J'({\bf k})<0$: $S(k,\omega)$ has a $\delta$-function peak
  below the continuous spectrum. A typical example is shown in
  Fig.~\ref{fig:skw}~(a).  As discussed above, this peak arises from a
  triplon bound state of two spinons, $|\Psi_{1{\bf k}}\rangle$.  The
  triplon dispersion $\omega_t({\bf k})$ is determined from 
  the pole of \eqref{eq:RPA} where
\begin{equation}
  1 + J'({\bf k}) \chi'_{1D}(k_x,\omega_t({\bf k})) = 0
  \label{eq:5a}
  \end{equation}
  and $\chi''_{1D}(k_x,\omega_t({\bf k})) = 0$ outside the continuum.
  The pole appears below $\omega_{2,l}$ because there $\chi'_{1D}$ is
  positive.  The inter-chain dispersion of the triplon is due to the
  $k_y$-dependence of $J'({\bf k})$.  In the weak interchain-coupling
  regime, the spectral weight $Z$ and binding energy $\delta E =
  \omega_{2,l}(k_x) - \omega_t({\bf k})$ of the peak are small, and
  behave as $Z\sim |J'(k_x,k_y)|$ and $\delta E \sim |J'(k_x,k_y)|^2$
  (up to logarithmic corrections).  See the Supplementary Material for
  details.
\item $J'({\bf k})>0$: The spectral weight shifts upwards, and the
  peak is broadened in the continuum, see Fig.~\ref{fig:skw}~(b).  A
  suppression of spectral weight at the lower edge of the continuum
  occurs due to repulsion between the two spinons.  When $J'(k_x,k_y)$
  is sufficiently large, a $\delta$-function peak appears {\sl above}
  the two-spinon continuum.  This peak corresponds to an anti-bound
  triplon state. However, the anti-bound peak is
  broadened by the four-spinon contribution, which leads to non-zero
  spectral density above the two-spinon upper-boundary,
  $\omega>\omega_{2,u}$ \cite{caux}.
\item $J'({\bf k})=0$: For such momenta, the structure factor is {\sl
     identical} in the two-spinon approximation to that of a set of
   decoupled chains.  For the frustrated situation of principle
   interest, where $J'_1=J'_2+J'_3$, this condition is always satisfied
   for $k_x=\pi$ (but it may also be true elsewhere).
   \end{enumerate}
\par
%{\it Comparison with experiments - }
Now, let us compare the above features with the experimental results
\cite{ColdeaPRL,ColdeaPRB} on Cs$_2$CuCl$_4$.  The coupling constants
are experimentally estimated as $J$=0.374(5) meV,
$J'=J'_1$=$J'_2$=0.128(5) meV, which leads to the ratio $J'/J$=0.34(3)
\cite{ColdeaModel}.  This compound also has some very weak additional
Dzyaloshinskii-Moriya and interplane interactions not included in our
model.  These have significant effects only at very low energies, e.g.
in inducing long-range order in the ground state and weak
incommensurability of the ordering wave vector 
\cite{RPA_Cs2CuCl4,triangDimer}.
The coupling constants of these interactions are experimentally
estimated as about 0.05$J$ \cite{ColdeaModel}.  In this paper, we
neglect them for simplicity and discuss the physics for energies
higher than about 0.1$J$ -- note that the majority of the features in
the neutron scattering data in Refs.~\cite{ColdeaPRL,ColdeaPRB} are in
this higher energy regime. In the notation of
Refs.~\cite{ColdeaPRL,ColdeaPRB}, the Fourier component of the
interchain couplings reads
$J'({\bf k}) =4J'\cos(k'_x/2)\cos(k'_y/2) 
$, where $k'_x$ and $k'_y$
are the momenta corresponding to $b$ and $c$ axes in Refs.
\cite{ColdeaPRL,ColdeaPRB}, respectively: $k'_x=k_x$ and $k'_y=k_x+2k_y$.
\par
\begin{figure} \includegraphics[scale=0.7]{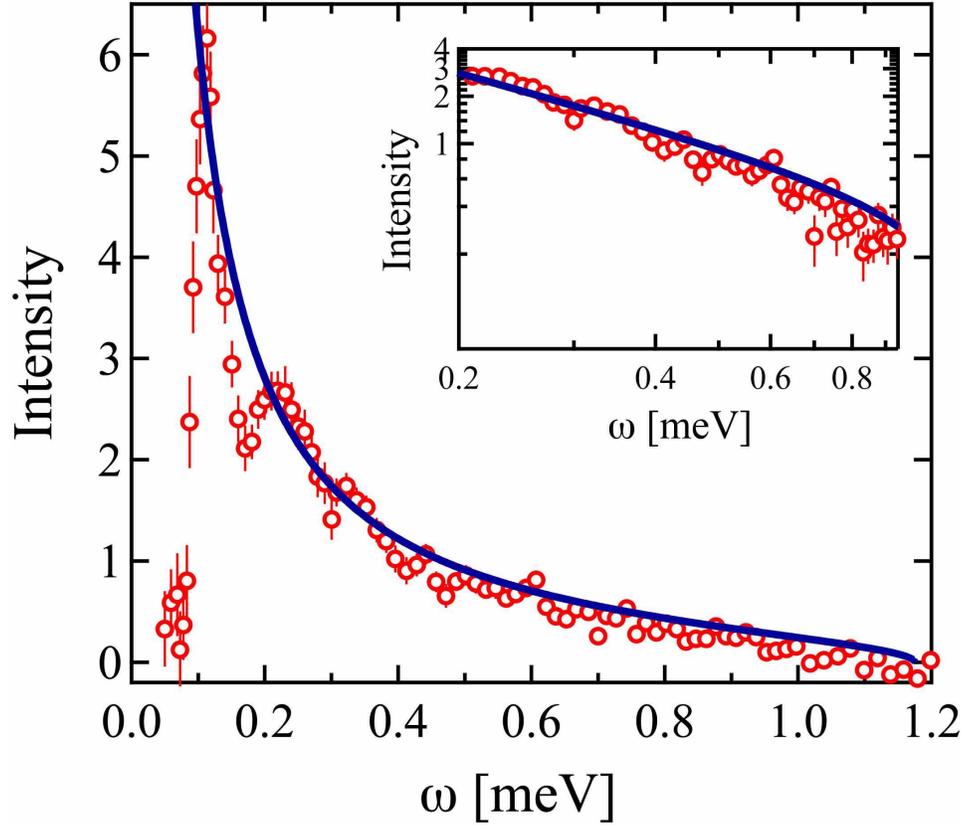} \caption{Comparison
    with the experimental result for dynamical structure factor
    $S({\bf k},\omega)$ at $k'_x=\pi$.  Solid line denotes the
    two-spinon structure factor $S_{1D}(\pi,\omega)$ of a single chain
    with exchange $J = 0.374$ meV\cite{ColdeaPRB}. The symbols are the
    experimental data obtained by the inelastic neutron scattering
    experiment on Cs$_2$CuCl$_4$, taken from the G scan of Fig. 5 in
    Ref. \cite{ColdeaPRB}. The inset shows the log-log plot. The
    theoretical result is fitted to the experimental data by adjusting
    the height with a single multiplication factor.  The experimental
    data in this and the following figures are excerpted with
    permission from Ref.~\cite{ColdeaPRB}.  Copyright (2003) by the
    American Physical Society.}
\label{fig:expskw} \end{figure}

First, we discuss the large tail of $S({\bf k},\omega)$ and the
interpretation of the power-law behaviors observed in Cs$_2$CuCl$_4$
\cite{ColdeaPRB}.  In the present approach, a power-law behavior at
the lower edge of the continuum ($\omega_{2,l}$) is obtained only when
$J'({\bf k})=0$. There,
we expect the same behavior as occurs in decoupled Heisenberg chains,
i.e.
$S({\bf k},\omega)\propto\sqrt{-\ln[\omega-\omega_{2,l}]/[\omega- 
\omega_{2,l}]}$
at $k_x\ne\pi$, and $S({\bf k},\omega)\propto\sqrt{-\ln\omega}/\omega 
$ at
$k_x=\pi$ near the lower edge of
continuum
\cite{chainMqwcalc}.  On a spatially anisotropic triangular lattice,
$J'({\bf k})$ is zero on the lines of $k_x=\pi$ and $k_y=(\pi- 
k_x)/2$
in momentum space, which correspond to the lines of $k'_x=\pi$ and
$k'_y=\pi$.  The experimental result at $k'_x=\pi$ is given as the G
scan in Ref. \cite{ColdeaPRB}.  The comparison of $S({\bf k},\omega)$ at
$k_x'=\pi$ between the present result (i.e. $S_{1D}(k,\omega)$ of the
Heisenberg chain) and the experimental data (G scan in Fig. 5 of Ref.
\cite{ColdeaPRB}) is shown in Fig. \ref{fig:expskw}.  Only a single
fitting parameter -- for the global height of intensity in this plot
-- has been employed.  For all further comparisons (below), we will
employ the same normalization, so the remaining comparisons are
parameter-free. Although the theoretical curve and experimental data
differ somewhat at low energies due to the neglect of long-range
magnetic order and the Dzyaloshinskii-Moriya interaction in the theory,
the agreement at higher energy is quite good.
\par
\begin{figure} \includegraphics[scale=1.0]{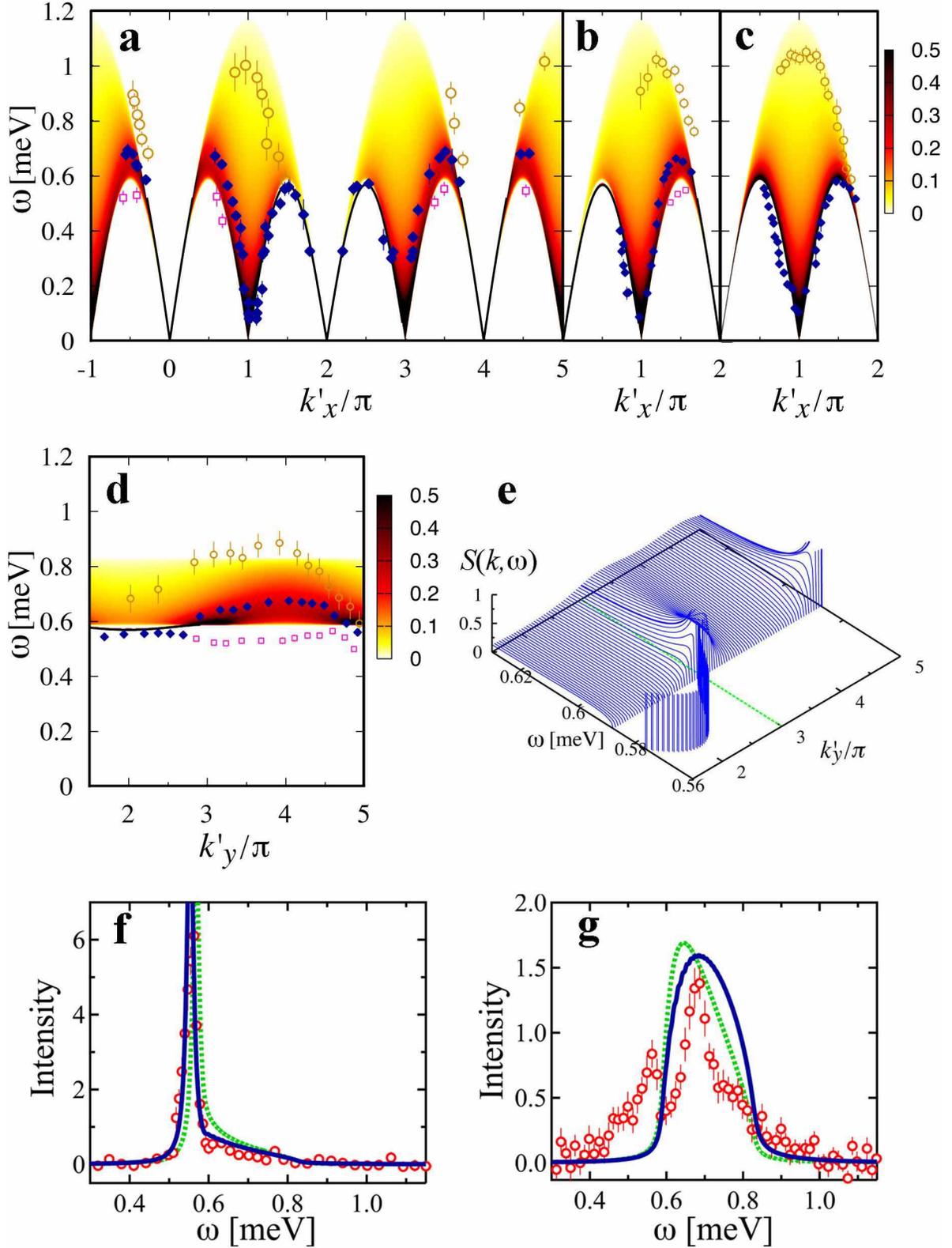} \caption{{\bf
a.}-{\bf d.} Comparison with experimental results for dispersion
relations at (a) $k'_y$=0, (b) $k'_y$=$2\pi$, (c) $k'_y$=3$\pi$ and
(d) $k'_x$=$-\pi/2$. Density plots are the present results of
dynamical structure factor $S({\bf k},\omega)$ for $J'_1=J'_2=0.34J$,
$J'_3=0$ and $J$=0.374 meV. Solid and open symbols denote the main
peak, and the upper and lower edges of the spectrum observed by the
neutron scattering experiment on Cs$_2$CuCl$_4$, respectively, taken
from Ref. \cite{ColdeaPRB}. Graphs (a)-(d) correspond to (1), (3), (4)
and (2) of Fig. 3 in Ref. \cite{ColdeaPRB}, respectively. {\bf e.}
$S({\bf k},\omega)$ at $k'_x=-\pi/2$ near the lower edge of continuum
obtained by the present approach. The sign of $J'({\bf k})$
changes at $k'_y=3\pi$. {\bf f.}-{\bf g.} Comparison with experimental
data for the line shape of $S({\bf k},\omega)$ at (f) ${\bf
k}'=(-\pi/2, 2\pi)$ and (g) ${\bf k}'=(-\pi/2, 4\pi)$. Dotted lines
are present results within the 2-spinon subspace multiplied by the
normalization factor obtained by fitting the G scan in
Fig.\ref{fig:expskw}. Solid lines are RPA result which accounts
for the 4-spinon states as obtained in a chain of length $L_x=288$,
see main text for the details. The numerical data in f and g are  
broadened by
energy resolution $\Delta E=0.019$ meV of the spectrometer
\cite{ColdeaPRB} including the isotropic magnetic form factor of
Cu$^{2+}$ ions. Symbols are experimental data for (f) E scan and (g) F
scan of Fig. 5 in Ref. \cite{ColdeaPRB}.}  \label{fig:expdisp}
\end{figure}

We next turn to the dispersion relation, which we define here, in
order to ease comparison with experimental data, by the location of
the peak $\omega({\bf k})$ in $S({\bf k},\omega)$ at each ${\bf k}$.
A comparison of our result and the experimental data (from Fig. 3 in
Ref.  \cite{ColdeaPRB}) is shown in Fig.~\ref{fig:expdisp}. It should
be noted that there is no fitting parameter in this plot.  The
asymmetry of the dispersion relation of the main peak with respect to
$k'_x=\pi$ and $3\pi$ observed at $k'_y$=0 and $2\pi$ is consistently
reproduced by the present approach (Fig.  \ref{fig:expdisp} (a,b)).
At $k'_y$=3$\pi$, the dispersion relation is symmetric because
$J'({\bf k})$ is zero at this momentum, which is also consistent with
the experimental observation (Fig.  \ref{fig:expdisp} (c)).  Despite
the 1D starting point of the approach, it explains the experimental
dependence upon transverse momentum  ($k'_y$) as well.
Figure \ref{fig:expdisp} (d,e) shows $S({\bf k},\omega)$ in the
perpendicular direction to $k'_x$ at $k'_x=-\pi/2$.  The sign of
$J'({\bf k})$ changes at $k'_y$=3$\pi$.  This causes the following
change in $S({\bf k},\omega)$: As shown in Fig.  \ref{fig:expdisp}
(e), a bound state is formed just below the continuum for $k'_y<3\pi$.
On the other hand, for $k'_y>3\pi$, the spectral weight shifts
upwards, and the peak is broadened and absorbed into the continuum.
Put simply, {\sl the lower edge of continuum (open squares in
  Fig.~\ref{fig:expdisp}~(a- d)) lies below the peak only in the
  region of $J'({\bf k})>0$, and the main peak is always observed at
  the lowest energy of the spectrum for $J'({\bf k})\le0$}.  These
features are exactly in accord with the theoretical predictions.
Moreover, for $J'({\bf k})<0$, the peak is much sharper (in fact
resolution limited) than for $J'({\bf k})>0$.  This is illustrated in
Figs.~\ref{fig:expdisp}~(f,g), which compare our theoretical
predictions to scans E,F of Ref.~\onlinecite{ColdeaPRB} -- note the
factor of $4$ larger scale in Fig.~\ref{fig:expdisp}~(f) compared to
Fig.~\ref{fig:expdisp}~(g).

Furthermore, the asymmetry of the experimental estimate of the upper
edge of continuum with respect to $k'_x=\pi$ or $k'_y=3\pi$ is also
qualitatively understood: At the momenta with
$J'({\bf k})>0$, the spectral weight shifts upwards, and the
high-energy weight becomes larger. On the other hand, in the region of
$J'({\bf k})<0$, the high-energy weight decreases, because part
of it shifts into the bound state (Figs. \ref{fig:skw} and
\ref{fig:expdisp} (e)). This feature is consistent with the behavior of
the upper edge of continuum observed in the experiment (open circles in
Fig. \ref{fig:expdisp} (a-d)). Namely, the peak of the dispersion
relation of the upper edge of continuum is observed at the momentum a
little shifted toward the region of $J'({\bf k})>0$ from
$k'_x=\pi$ or $k'_y=3\pi$.

Our approach allows for systematic improvements by including further
multi-spinon states.  As a first step, we included the four-spinon states in
the RPA approximation.  This is done numerically by expressing the
matrix element in Eq.\eqref{eq:9} for a finite length Heisenberg chain
as a product of determinants\cite{DetBethe1,DetBethe2,DetBethe3}. The
sum rule for the total spectral weight and the first frequency moment is
satisfied by more than 99\% for the length ($L_x=288$) considered.  We
then calculate from this Eq.\eqref{eq:RPA} using $\chi'_{1D} 
\rightarrow{\rm Re}\chi_{1D}$ and $\chi_{1D}''\rightarrow{\rm Im}\chi_ 
{1D}$ and obtain the
two-dimensional $S({\bf k}, \omega)$.  We note
that the finite-size errors for $L_x=288$ are insignificant compared to
the instrumental resolution.  The resulting changes are small but very
encouraging -- the bound state in scan E has moved down a little, making
agreement with experimental data essentially perfect (see
Fig.~\ref{fig:expdisp}~(f)).  We also observe that the anti-bound
states, being located in the region of $\omega - {\bf k}$ space with
non-zero spectral weight for 4-spinon excitations, acquire a non-zero
linewidth as expected, but that this is small enough that they remain
visible features.

We conclude with a general discussion of our method and its
ramifications.  The most significant feature
is the emergence of a spinon bound state driven by kinetic
energy.  Despite the superficial similarity to the more familiar {\sl
   magnon}, the physics of the bound state is quite distinct.
Specifically, a magnon is a Goldstone mode which emerges in a
long-range ordered magnet as a consequence of broken symmetry. In our
calculations, no such broken symmetry is presumed.  Instead, the bound
state is a true $s=1$ triplet excitation, and is better characterized
as a {\sl triplon} than a magnon.  The same is true for the anti-bound
state.  In fact, the anti-bound triplon is directly analogous to the
{\sl zero sound} mode of a neutral interacting Fermi gas\cite{agd}.

Since in most cases, weakly coupled spin chains {\sl do} eventually
order at low enough temperature, it is important to understand the
validity of our scheme in this situation.  For this, it is crucial
that we consider {\sl frustrated} inter-chain couplings
($J'_1=J'_2+J'_3$).  In this case, the leading divergence associated
with coupling neighboring chains -- the strong tendency to N\'eel
order at $k_x=\pi$ within each chain -- is removed because
$J'(\pi,k_y)=0$.  Without this condition, one
obtains\cite{RPAq1D_Scalapino,RPAq1D_Schulz} strong long-range N\'eel
order which influences spectral features on the scale of $O(J')$.
Since this effect is comparable to those captured by the two-spinon
approximation, the latter is unjustified without frustration.  With
frustration, any fluctuation-induced order has a much smaller
characteristic energy scale\cite{J1J2Dimer,triangDimer,RPA_Cs2CuCl4},
and can be neglected compared to the shifts of excited states captured
by the present approach. Of course, the presence of {\sl any}
long-range order, however weak, does modify some excitations in a
qualitative manner.  The triplon, when present, is expected to
transform smoothly into a magnon as a consequence.  In regions of
momentum space where no bound state is present below the continuum,
$J'({\bf k})>0$, a magnon may weakly emerge as a consequence of
long-range order.

There are numerous important directions for extensions and
applications.  It would be interesting to compare with neutron
measurements of Cs$_2$CuBr$_4$, which is isostructural and can be
modeled similarly to Cs$_2$CuCl$_4$ but with somewhat larger $J'/J
\approx 0.5$\cite{CsCuBr}, and to search for signs of the anti-bound
triplon in either material.  Some  theoretical extensions
would be to include three-dimensional and Dyzaloshinskii-Moriya
couplings, systematically treat higher-spinon states, to include
thermal fluctuations at $T>0$, and to take into account weak
long-range order.  A very interesting different direction is to apply
analogous methods to spatially anisotropic strongly interacting {\sl
  conductors}, modeled by Hubbard or $t$-$J$ type Hamiltonians.  Given
the very small arsenal of theoretical techniques capable of reliably
obtaining intermediate energy spectra in strongly interacting systems
above one dimension, further investigation of such methodology seems
highly worthwhile.
%{\it Acknowlegements - }
We would like to thank J. Alicea, M.P.A. Fisher and R. Shindou for
discussions.  This work is supported by the Grant-in-aid for Scientific
Research (C) No. 10354143 from MEXT, Japan (M.~K.), the Petroleum
Research Fund ACS PRF 43219-AC10 (O.~S.), NSF grant/DMR-0457440 (L.~B.)
and the Packard Foundation (L.~B.).  Part of this research was completed
at KITP and supported in part by NSF under Grant No. PHY05-51164.

\section*{Supplementary Material}
\subsection*{Basis}

The two-spinon states of a single chain are characterized by two
continuous quantum numbers, which can be thought of either as the
momenta $k_{x1},k_{x2}$ of the individual (unbound) spinons, or
equivalently, the total momentum $k_x=k_{x1}+k_{x2}$ and (excitation)
energy $\epsilon=\epsilon_s(k_{x1})+\epsilon_s(k_{x2})$ of the pair.
The spinon excitation energy is given by the des Cloizeaux-Pearson
dispersion relation $\epsilon_s(k) = (\pi J/2) \sin[k]$ \cite{dCP},
which is seen to describe the lower boundary of the two-spinon
continuum, $\epsilon_s(k) = \omega_{2,l}(k)$.  The excitation energy of
the spinon pair is then expressed via the total, $k_x$, and relative,  
$q_x =
(k_{x1}-k_{x2})/2$, momenta of the pair $\epsilon(k_x,q_x) = \pi J
\sin[k_x/2] \cos[q_x]$. Observe that the upper (lower) boundaries of the
two-spinon continuum correspond to $q_x=0$ ($\pm k_x/2$).  We find it
convenient to describe the two-spinon state of a chain in terms of the
total momentum $k_x$ and excitation energy $\epsilon$ of the pair.  The
transformation from $q_x$ to $\epsilon$ explains the density of states
factor $D_{k_x}(\epsilon)$ appearing in (\ref{eq:psi}).

To derive (\ref{eq:2}), we evaluate the expectation value of the
Hamiltonian (\ref{eq:1}) in the state (\ref{eq:psi})
\begin{equation}
\langle \Psi| {\cal H} |\Psi\rangle_{\bf k} = E_0 L_x L_y + \int\! d  
\epsilon D_{k_x}(\epsilon) \epsilon |\psi_{k_x}(\epsilon)|^2
+ J'({\bf k}) \int\! d \epsilon d \tilde{\epsilon} D_{k_x}(\epsilon)  
D_{k_x}(\tilde{\epsilon}) A_{k_x}^{*}(\epsilon) A_{k_x}(\tilde 
{\epsilon})
\psi^*_{k_x}(\epsilon) \psi_{k_x}(\tilde{\epsilon}),
\label{eq:mat-element}
\end{equation}
where $E_0 = J(-\ln2+1/4)$ is the ground state energy per site of  
decoupled chains \cite{BetheAnsatz,1DHeisE},
and we made use of the normalization condition $\int\! d \epsilon D_ 
{k_x}(\epsilon)|\psi_{k_x}(\epsilon)|^2 = 1$
as appropriate for the state (\ref{eq:psi}). ``Factoring out'' $ 
\epsilon$-integration $\int d\epsilon D_{k_x}(\epsilon) \psi^*_{k_x} 
(\epsilon)$
in (\ref{eq:mat-element}) leads to Eq.(\ref{eq:2}) of the main text.

The ground state to two spinon matrix element $A_{k_x}(\epsilon)$
represents the key technical element of our calculation
\begin{equation}
A_{k_x}(\epsilon)\equiv
\sqrt{2}\langle0|S^x_{-k_x,y}|k_x,\epsilon\rangle_y=-i\sqrt{2} 
\langle0|S^y_{-k_x,y}|k_x,\epsilon\rangle_y .
\label{eq:A1}
\end{equation}
Its absolute value squared, $M(k_x,\epsilon)=|A_{k_x}(\epsilon)|^2$,
also called the singlet-to-triplet transition rate, is obtained exactly
by an algebraic analysis based on infinite-dimensional quantum group
symmetries in Ref. \cite{chainMqw} and subsequently simplified in Ref.
\cite{chainMqwcalc}, which we follow here.  The matrix element $A$ in
(\ref{eq:A1}) is obtained as a square-root of $M$ because two-spinon
states with different $k_x$ and/or $\epsilon$ are orthogonal and thus
the phase can be set to zero for every set $(k_x,\epsilon)$
independently.  Note that our choice of $|k_x,\epsilon_{k_x}\rangle_y$
as an $S^z=+1$ eigenstate of 2 spinons in $y$-th chain ensures that
$\langle0|S^z_{-k_x,y}|k_x,\epsilon\rangle_y=0$.

Specifically, $A_k(\epsilon) = \exp[- I(t)/2]/\sqrt{4\pi}$, where  
\cite{chainMqwcalc}
\begin{equation}
I(t) = - I_0 - \ln| t \sinh(\pi t/4)| + s(t) ~,~ s(t) = \int_0^\infty  
dx \frac{\sin^2[xt/2]}{x \cosh^2[x]} ,
\end{equation}
$I_0 = 0.3677...$ and $k,\epsilon$ dependence comes via the parameter  
$t$
\begin{equation}
\cosh[ \frac{\pi t}{4}] = \sqrt{\frac{\omega_{2,u}^2(k) - \omega_{2,l} 
^2(k)}{\epsilon^2 - \omega_{2,l}^2(k)}} .
\end{equation}

Integration of $s(t)$ has been performed numerically by an adaptive  
quadrature algorithm.

\subsection*{Discretization in the $\epsilon$ space}
For numerical calculations, we carefully discretize energy $\epsilon$  
for every
$k_x$ value, by dividing the interval $\omega_{2,u}(k_x) - \omega_ 
{2,l}(k_x)$ into
$M$ discrete points.
The resulting discrete $M\times M$ eigenvalue problem is then solved
for every value of momentum  ${\bf k} = (k_x,k_y)$.
The data points in the $\epsilon$-space
are chosen so that the distribution of them reduces to the exact  
density of states of the Heisenberg
chain \cite{2spinonE} in the continuous limit:
\begin{equation}
\Delta\epsilon D_{k_x}(\epsilon)=\frac{|k_x|}{2M} .
\label{eq:DOS}
\end{equation}
Next, it is convenient to define rescaled
two-spinon states with Kronecker delta function overlaps in place of the
Dirac delta-function overlapping continuum states.  Specifically:
\begin{equation}
   \label{eq:8}
   |\epsilon\rangle^M \equiv \sqrt{\frac{|k_x|}{2M}} |\epsilon\rangle.
\end{equation}
One can check that
\begin{equation}
   \label{eq:13}
   {}^M\langle \epsilon'|\epsilon\rangle^M = \frac{k_x}{2M} \langle
   \epsilon'|\epsilon\rangle = \frac{k_x}{2M} \frac{1}{D(\epsilon)}
   \delta(\epsilon-\epsilon') = \frac{k_x}{2M}
   \frac{1}{D(\epsilon)\Delta\epsilon} \delta_{\epsilon,\epsilon'} =
   \delta_{\epsilon,\epsilon'}
\end{equation}
as desired. We also rescale wavefunction as
  $\phi(\epsilon) = \sqrt{|k_x|/(2M)} \psi_{\bf k}(\epsilon)$,
suppressing for brevity its dependence on the center-of-mass momentum
${\bf k}$ (it enters the problem only as a parameter).
With these definitions, equation \eqref{eq:psi} takes on simple form
\begin{equation}
|\Psi_{\bf k}\rangle = \sum_\epsilon \phi(\epsilon) |\epsilon  
\rangle^M .
\label{eq:psi-M}
\end{equation}
The Schr\"odinger equation, in turn, takes on matrix form
\begin{equation}
  \epsilon \phi(\epsilon) + J'({\bf k}) \frac{|k_x|}{2M}  \sum_{\tilde 
{\epsilon}}
  A^*(\tilde{\epsilon}) A(\epsilon) \phi(\tilde{\epsilon}) = E_{\bf  
k} \phi(\epsilon) .
  \label{eq:SE-M}
  \end{equation}
We typically took the number of discretized energies $M$ to be  
several thousand,
as large as necessary to ensure good resolution and convergence of  
the results.

\subsection*{Dynamical structure factor $S(k,\omega)$}
The dynamical structure factor $S({\bf k},\omega)$ is defined by
\begin{equation}
S({\bf k},\omega)\equiv\int\frac{dt}{2\pi}{\rm e}^{{\rm i}\omega t} 
\langle
{\rm G.S.}|
S^\alpha_{-{\bf k}}(t)S^\alpha_{\bf k}(0)|{\rm G.S.}\rangle,
\label{eq:Skw1}
\end{equation}
where {\sl no sum} on $\alpha=x,y$ or $z$ is implied.  This
is calculated within the 2-spinon subspace using the obtained
eigenstates and energies of the effective Hamiltonian as
\begin{equation}
S({\bf k},\omega)=\frac{1}{2}\sum_{E_{\bf k}} |\langle {\rm
   G.S.}|S^-_{-{\bf k}}|\Psi_{\bf
   k}\rangle|^2
\delta(\omega-E_{\bf k}),
\label{eq:Skw3}
\end{equation}
where summation is over excited states of the system with momentum ${\bf
   k}$ and energy $\omega=E_{\bf k}$.
In Eq.\eqref{eq:Skw3}, it is clear that, since the ground state is a
singlet, only triplet excited states $|\Psi_{\bf k}\rangle$ with  
total $S^z=+1$
can contribute to the sum.

Note that
\begin{equation}
   \label{eq:22}
   |\Psi_{\bf k}\rangle = \sum_\epsilon \frac{1}{\sqrt{L_y}}\sum_y  
\phi(\epsilon) e^{ik_y
     y}|k_x,\epsilon,y\rangle^M ,
\end{equation}
and writing $S^-_{-{\bf k}} = \frac{1}{\sqrt{L_y}} \sum_{y'} e^{-ik_y
   y'} S^-_{-k_x,y'}$, we have
\begin{equation}
   \label{eq:23}
   S^-_{-{\bf k}}|\Psi_{\bf k}\rangle = \sum_\epsilon \frac{1}{L_y} 
\sum_{y,y'}
   \phi(\epsilon) e^{ik_y (y-y')}
   S^-_{-k_x,y'} |k_x,\epsilon,y\rangle^M .
\end{equation}
This can be separated into terms with $y'=y$ and $y'\neq y$.  Projecting
the resulting state into the subspace containing only zero or two
spinons per chain, one then obtains
\begin{equation}
   \label{eq:25}
   S^-_{-{\bf k}}|\Psi_{\bf k}\rangle = |\Upsilon\rangle_0 + |\Upsilon 
\rangle_1,
\end{equation}
with components containing zero or four total spinon excitations:
\begin{eqnarray}
   \label{eq:24}
   |\Upsilon\rangle_0 & = & \sqrt{2} \sqrt{\frac{|k_x|}{2M}} \sum_ 
\epsilon
   \phi(\epsilon)A_{k_x}(\epsilon) |{\rm G.S.}\rangle_0, \\
  |\Upsilon\rangle_1 & = &   \sum_{\epsilon,\epsilon'}
   \sqrt{2} \sqrt{\frac{|k_x|}{2M}}\frac{1}{L_y}\sum_{y\neq y'} e^ 
{ik_y (y-y')}
   \phi(\epsilon)
   A_{-k_x}^*(\epsilon')|k_x,\epsilon;S^z=+1\rangle^M_y |-k_x, 
\epsilon';S^z=-1\rangle^M _{y'}
   \otimes_{y''\neq y,y'} |0\rangle_{y''}.\label{eq:27}
\end{eqnarray}
Note that in Eq.\eqref{eq:27} we have indicated explicitly the total
$S^z$ of the spinon pair on chains $y,y'$, since the two chains have
equal and opposite $S^z=\pm 1$.

We approximate the ground state by its form to first order in $J'({\bf
   k})$, in the subspace of states with zero or two spinons per chain:
\begin{equation}
   \label{eq:14}
   |{\rm G.S.}\rangle \approx |{\rm G.S.}\rangle_0 + |{\rm G.S.} 
\rangle_1,
\end{equation}
with as usual, in first order perturbation theory,
\begin{equation}
   \label{eq:21}
   |{\rm G.S.}\rangle_1 = \frac{1}{E_0-{\mathcal H}_0} {\mathcal H}'
   |{\rm G.S.}\rangle_0.
\end{equation}
Here ${\mathcal H}_0= \left.{\mathcal H}\right|_{J'_a\rightarrow0}$  
is the
decoupled chain Hamiltonian, and ${\mathcal H}'={\mathcal H}-{\mathcal
   H}_0$ contains the interchain exchange couplings.

The desired matrix element then has two terms:
\begin{equation}
   \label{eq:26}
   \langle
{\rm G.S.}|
S^-_{-{\bf k}}|\Psi_{\bf k}\rangle = {}_0\langle
   {\rm G.S.}| \Upsilon\rangle_0 +  {}_1\langle
   {\rm G.S.}| \Upsilon\rangle_1 .
\end{equation}

The first term is simplest. From Eq.\eqref{eq:24}, one directly obtains
\begin{equation}
   \label{eq:28}
   {}_0\langle
   {\rm G.S.}| \Upsilon\rangle_0 = \sqrt{2} \sqrt{\frac{|k_x|}{2M}}  
\sum_\epsilon
   \phi(\epsilon)A_{k_x}(\epsilon).
\end{equation}
Now consider the second term.   To evaluate this explicitly, it is  
useful to write
\begin{equation}
   \label{eq:29}
  {\mathcal H}' = \sum_{k_x,y} J'(k_x)\left[ \frac{1}{2} \left(S^+_ 
{k_x,y}
    S^-_{-k_x,y+1} + S^-_{k_x,y}
    S^+_{-k_x,y+1} \right) + S^z_{k_x,y} S^z_{-k_x,y+1}\right].
\end{equation}
Because $|\Upsilon\rangle_1$ contains
only spin $S^z=\pm 1$ spinon pairs on two chains, we can restrict the
consideration of $|{\rm G.S.}\rangle_1$ to those components with the
same structure.  This means that the third term inside the square
brackets in Eq.\eqref{eq:29} can be neglected, since it creates $S^z=0$
two-spinon states on the chains $y,y+1$.  Taking only the first two
terms, we have
\begin{equation}
   \label{eq:30}
   |{\rm G.S.}\rangle_1 = -\sum_{k_x,y}
   \frac{|k_x|}{2M}J'(k_x)\sum_{\epsilon,\epsilon'} \frac{A_{k_x}^* 
(\epsilon)
   A_{-k_x}^*(\epsilon')}{\epsilon+\epsilon'} \sum_{\delta=\pm 1} | 
k_x,\epsilon;S^z=\delta\rangle^M_y |-k_x,\epsilon';S^z=-\delta 
\rangle^M _{y+1}
   \otimes_{y''\neq y,y+1} |0\rangle_{y''}.
\end{equation}
We can then evaluate the overlap.  One obtains
\begin{eqnarray}
   \label{eq:32}
   {}_1\langle
   {\rm G.S.}| \Upsilon\rangle_1 & = & -\sqrt{2} \left(
     \frac{|k_x|}{2M}\right)^{3/2} J'({\bf k})\sum_{\epsilon,\epsilon'}
   \frac{\phi(\epsilon) A_{k_x}(\epsilon) A_{-k_x}^*(\epsilon')
     A_{-k_x}(\epsilon')}{\epsilon+\epsilon'} .
\end{eqnarray}
Observe that the correction diverges at
$k_x=\pi$, where the spinons have zero energy, {\sl unless}
$J'(k_x=\pi)=0$.  This
hints at the instability of the system towards magnetic ordering in
the case of {\sl non-frustrated} inter-chain coupling.

\begin{figure} \includegraphics[scale=0.35]{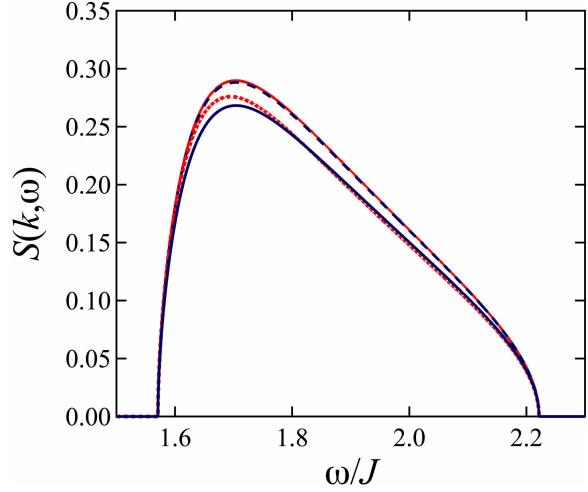} \caption{Comparison
of $S({\bf k},\omega)$ at ${\bf k}'=(-\pi/2, 4\pi)$ between the
present approximation \eqref{eq:Skw3} and \eqref{eq:Skw4} 
(blue solid line), that without the correction
to the ground state (blue dashed line), RPA as derived in \eqref{eq:9},
\eqref{eq:Scontfinal} and \eqref{eq:chi} (red line), and 
the standard RPA \eqref{eq:standard-RPA} (dotted red line).}
\label{fig:Skw-RPA}
\end{figure}

Combining Eqs.(\ref{eq:32},\ref{eq:28}), the perturbation-theory- 
improved
transition rate can be expressed as
\begin{equation}
|\langle
{\rm G.S.}|
S^-_{-{\bf k}}|\Psi_{\bf k}\rangle|^2 = \frac{|k_x|}{M}
   \Biggl|\sum_\epsilon
   \phi(\epsilon)A_{k_x}(\epsilon)\left\{1-J'({\bf k})\sum_{\epsilon'} 
\frac{|k_x|}{2M}
     \frac{|A_{k_x}(\epsilon')|^2}{\epsilon+\epsilon'}\right\}\Biggr|^2.
\label{eq:Skw4}
\end{equation}
This and equation \eqref{eq:Skw3} provides way to numerically evaluate
the structure factor. Fig. \ref{fig:Skw-RPA} shows that effect of the
perturbative correction ${}_1\langle {\rm G.S.}| \Upsilon\rangle_1$
on the calculated structure factor is very small.

\subsection*{Connection to random-phase approximation (RPA)}

It is possible to obtain an analytic solution to the integral equation
describing the two-spinon states, Eq.\eqref{eq:2}.  To do so, it is
convenient to use the same
discretization scheme as described above in Eq.\eqref{eq:DOS}
to resolve the wavefunctions of individual states in the
continuum.  The Schr\"odinger equation \eqref{eq:SE-M} is re-written as
\begin{equation}
   \label{eq:3}
   \epsilon \phi(\epsilon) + B(E) A(\epsilon) = E\phi(\epsilon).
\end{equation}
Note that the quantity
\begin{equation}
   \label{eq:4}
   B(E)= \frac{J'({\bf k})|k_x|}{2M}\sum_{\tilde\epsilon}
   \phi(\tilde\epsilon)A^*(\tilde\epsilon)
\end{equation}
is {\sl independent} of $\epsilon$.  This allows one to completely
determine the $\epsilon$ dependence of $\phi(\epsilon)$ as
\begin{equation}
   \label{eq:5}
   \phi_E(\epsilon) = \frac{B(E) A(\epsilon)}{E-\epsilon}.
\end{equation}
Inserting this form into Eq.\eqref{eq:4}, one obtains the eigenvalue
condition
\begin{equation}
   \label{eq:6}
   \frac{J'({\bf
       k})|k_x|}{2M}\sum_{\epsilon}
   \frac{|A(\epsilon)|^2}{E-\epsilon} = 1.
\end{equation}
This equation has two classes of solutions.  There are bound (or
anti-bound) states, in which the eigenvalue $E$ is well-separated from
the continuum of the decoupled chains.  Then the denominator in
Eq.\eqref{eq:6} remains finite as one takes the discretization to zero,
i.e. $M\rightarrow\infty$.  On thereby obtains the bound state condition
\begin{equation}
   \label{eq:7}
  J'({\bf
       k}) \int\! d\epsilon\, D(\epsilon) \frac{|A(\epsilon)|^2}{E- 
\epsilon} = 1.
\end{equation}
The second class of solution is more subtle, and occurs when $E$ is in
the range of the continuum, i.e. for finite but large $M$, it is close
to one of the discretized two-spinon eigenvalues of the decoupled
chains, which we denote $\epsilon_0$.  We assume (and confirm
self-consistently) that the energy of such a state can be written as
$E=\epsilon_0 + \delta/M$, where $\delta$ remains $O(1)$ as
$M\rightarrow \infty$.  We then rewrite Eq.\eqref{eq:6} as
\begin{equation}
   \label{eq:10}
   \frac{J'({\bf
       k})|k_x|}{2M}\sum_{\epsilon} |A(\epsilon)|^2 \left[
     \frac{\epsilon_0-\epsilon}{(\epsilon_0-\epsilon)^2-
       (\delta/M)^2} -
     \frac{\delta/M}{(\epsilon_0-\epsilon)^2-(\delta/M)^2}\right] = 1.
\end{equation}
The first term contains $\epsilon_0-\epsilon$ in the {\sl numerator},
and the summand is locally {\sl odd} about $\epsilon=\epsilon_0$.  The
contribution from the sum in the region where $|\epsilon-\epsilon_0|$ is
$O(1/M)$ is therefore negligible, because the contributions from
eigenvalues $\epsilon$ on either side of $\epsilon_0$ cancel.
Conversely, there is a non-vanishing contribution from
$|\epsilon-\epsilon_0|$ of $O(1)$, which in the $M\rightarrow \infty$
limit becomes a principle part integral.  Conversely, in the second sum,
there is only a $\delta/M$ factor in the numerator, and the integrand is
locally {\sl even} about $\epsilon=\epsilon_0$.  In this sum, there is
an $O(1)$ contribution from the region $|\epsilon-\epsilon_0|$ of
$O(1/M)$.  This must be calculated explicitly, by summing over discrete
$\epsilon_n=\epsilon_0 + \Delta\epsilon n$, with integer $n$. Here the
level spacing is determined from $D(\epsilon_0)\Delta \epsilon =
|k_x|/(2M)$, and because the sum is sharply peaked we can approximate
$A(\epsilon) \approx A(\epsilon_0)$.  Because the integrand decays as
$|\epsilon-\epsilon_0|^{-2}$, the contribution from
$|\epsilon-\epsilon_0|$ of $O(1)$ is negligible in the sum, and the
limits on $n$ can be extended to $\pm \infty$.

Carrying out this sum (using $2b^2\sum_1^\infty 1/(n^2 -b^2) = 1 -  
\pi b \cot(\pi b)$)
and taking the $M\rightarrow \infty$ limit, we
find the simple result
\begin{equation}
   \label{eq:11}
   \pi J'({\bf k}) D(\epsilon_0) |A(\epsilon_0)|^2 \cot (\pi \gamma) +
   J'({\bf k}) P\int\!
   d\epsilon \, \frac{D(\epsilon) |A(\epsilon)|^2}{\epsilon_0- 
\epsilon} =
   1,
\end{equation}
where we have defined for convenience
\begin{equation}
   \label{eq:12}
   \gamma = \frac{2D(\epsilon_0) \delta}{|k_x|} = \frac{\delta E}{\Delta
     \epsilon}.
\end{equation}
This form guarantees $|\gamma| < 1/2$, so that the shift of the energy
level $\delta E$ is always less than half the distance to the nearest
eigenvalue, i.e. the levels do not cross upon increasing $J'$.

Now we turn to the determination of the structure factor.
Normalization of state \eqref{eq:psi-M} fixes $|B(E)|$ according to
\begin{equation}
   \label{eq:15}
   |B(E)|^2 = \left[\sum_\epsilon
     \frac{|A(\epsilon)|^2}{(E-\epsilon)^2}\right]^{-1} .
\end{equation}
This observation  leads us to the structure factor, which we
divide into the bound state and continuum contributions:
\begin{equation}
   \label{eq:20}
  S({\bf k},\omega) =   S_{\rm bs}({\bf k},\omega)+ S_{\rm cont}({\bf  
k},\omega).
\end{equation}
First
consider the bound (anti-bound) state contribution -- we will consider
only a single bound (anti-bound) state, since this occurs for the
triangular lattice of present interest.  In any case, multiple states  
would
simply give additive contributions.  This gives directly
a delta-function peak in the structure factor, at $\omega=E_{bs}$, where
$E_{bs}$ is the bound state energy:
\begin{equation}
   \label{eq:16}
  S_{\rm bs}({\bf k},\omega)=|B(E)|^2  \frac{|k_x|}{2M}\sum_{\epsilon, 
\epsilon'}
  \frac{|A(\epsilon)|^2|A(\epsilon')|^2}{(E-\epsilon)(E-\epsilon')} 
\delta(\omega-E) .
\end{equation}
Using Eq.\eqref{eq:6}, this immediately simplifies to
\begin{equation}
   \label{eq:17}
   S_{\rm bs}({\bf k},\omega)=|B(E)|^2 \frac{2M}{|k_x|} \frac{1} 
{[J'({\bf k})]^2}\delta(\omega-E).
\end{equation}
In this case, since $E-\epsilon$ remains non-zero as $M\rightarrow
\infty$, the sum in Eq.\eqref{eq:15} can be converted to an integral:
\begin{equation}
   \label{eq:18}
   |B(E)|^2 =  \frac{|k_x|}{2M} \left[\int\! d\epsilon\,D(\epsilon)
     \frac{|A(\epsilon)|^2}{(E-\epsilon)^2}\right]^{-1} .
\end{equation}
Thus we obtain the bound state delta-function contribution
\begin{equation}
   \label{eq:19}
   S_{\rm bs}({\bf k},\omega) = \left\{[J'({\bf k})]^2 \int\! d 
\epsilon\,D(\epsilon)
     \frac{|A(\epsilon)|^2}{(E_{bs}-\epsilon)^2}\right\}^{-1}
   \delta(\omega-E_{bs}).
\end{equation}

We now turn to the continuum contribution.  Following the same steps
as above, we find the analog of Eq.\eqref{eq:17},
\begin{equation}
   \label{eq:basicSform}
   S_{\rm cont}({\bf k},\omega)=\sum_{E\neq E_{bs}} |B(E)|^2 \frac{2M} 
{|k_x|} \frac{1}{[J'({\bf k})]^2}\delta(\omega-E).
\end{equation}
In this case, more care must be taken in evaluating $B(E)$ from
Eq.\eqref{eq:15}, because the energy denominators in the sum become
small as $M\rightarrow \infty$.  Indeed, the sum is dominated by
$|E-\epsilon|$ of $O(1/M)$, and so one may as in Eq.\eqref{eq:11}
consider the density of states (i.e. spacing $\Delta\epsilon$) and
$A(\epsilon)$ to be approximately constant in this region.  This
allows one to carry out the sum explicitly
(using $\sum_{-\infty}^\infty 1/(n+a)^2 = \pi^2/\sin^2(\pi a)$) and  
obtain
\begin{equation}
   \label{eq:BEcontinuum}
   |B(E)|^2= \frac{(\Delta\epsilon)^2}{\pi^2 \csc^2(\pi\gamma)}\frac 
{1}{|A(E)|^2}.
\end{equation}
Inserting this into Eq.\eqref{eq:basicSform}, in the large $M$ limit
the sum may be converted to an integral via $\sum_{E\neq E_{bs}}\Delta
E\rightarrow \int\! dE$, and thereby collapse the delta-function.  One
then obtains
\begin{equation}
   \label{eq:ContS}
    S_{\rm cont}({\bf k},\omega)=
    \frac{1}{\pi^2
      \csc^2(\pi\gamma)}\frac{1}{D(\omega)|A(\omega)|^2}\frac{1} 
{[J'({\bf
        k})]^2}.
\end{equation}
Using $\csc^2(\pi\gamma)=\cot^2(\pi\gamma)+1$ and Eq.\eqref{eq:11}, we
finally arrive at
\begin{equation}
   \label{eq:Scontfinal}
    S_{\rm cont}({\bf k},\omega)= \frac{S_{1D}(k_x,\omega)}{[1 + J'({\bf
        k}) \chi'(k_x,\omega)]^2
+ [J'({\bf k}) \chi''(k_x,\omega)]^2}
\end{equation}
with
\begin{equation}
\label{eq:chi}
\chi'(k_x,\omega)=
P \int_0^\infty\!
d\omega'\,\frac{D(\omega')|A(\omega')|^2}
{\omega'-\omega}
~ {\rm and}~
\frac{1}{\pi}\chi''(k_x,\omega)=S_{1D}(k_x,\omega)=D(\omega)|A 
(\omega)|^2 .
\end{equation}
The standard RPA result has the same functional form as \eqref 
{eq:Scontfinal} but with
$\chi',\chi''$ replaced by ${\rm Re}\chi_{\rm rpa}, {\rm Im}\chi_{\rm  
rpa}$ via
\begin{equation}
\label{eq:standard-RPA}
{\rm Re}\chi_{\rm rpa}(k_x,\omega)=
\frac{1}{\pi}P \int_{-\infty}^\infty\!d\omega'\,\frac{{\rm Im}\chi_{\rm rpa}(k_x,\omega')} 
{\omega'-\omega}
~ {\rm and}~
\frac{1}{\pi}{\rm {Im}}\chi_{\rm rpa}(k_x,\omega) = {\rm sgn}(\omega)  
S_{1D}(k_x,|\omega|) .
\end{equation}
The close similarity of expressions \eqref{eq:chi} and
\eqref{eq:standard-RPA} is illustrated in Fig.\ref{fig:Skw-RPA}.

\subsection*{Triplon dispersion in anisotropic triangular lattice}

Bound (and antibound) states outside the continuum are determined by the
condition $1 + J'({\bf k}) \chi'(k_x,\omega_t({\bf k})) =0$,
which is just \eqref{eq:7}, where $J'({\bf k}) = 4 J' \cos[k_x/2] \cos 
[k_y/2]$
for Cs$_2$CuCl$_4$.

{\sl Bound state:} Since $\chi' > 0$ for $\omega < \omega_{2,l}$ as  
follows from
\eqref{eq:chi}, one needs $J'(k_x,k_y) < 0$ for it to appear. Using  
the asymptotic
behavior of the chain structure factor near the lower edge of the  
two-spinon continuum,
$\omega_{2,l}$ \cite{chainMqwcalc}
$S_{1D} \approx C \sqrt{-\ln[\omega-\omega_{2,l}]/[\omega_{2,l} 
(\omega-\omega_{2,l})]}$,
we find with logarithmic accuracy
\begin{equation}
\label{eq:chiB}
\chi'(k_x,\omega \to \omega_{2,l}) = \sqrt{8} C \frac{\sqrt{-\ln 
[\omega_{2,l}-\omega]}}
{\sqrt{\omega_{2,l}[\omega_{2,l}-\omega]}} \arctan\sqrt{\frac{\omega_ 
{2,u}-\omega_{2,l}}
{\omega_{2,l}-\omega}} .
\end{equation}
Here $C = \exp[I_0/2]/\sqrt{16\pi^3}$. Provided that $J'({\bf k})$ is
negative, we readily see that the triplon binding energy behaves as
$\delta E = \omega_{2,l} - \omega_t({\bf k}) \propto [J'({\bf k})]^2
$, up to very weak logarithmic corrections. The triplon appears below
the continuum and propagates along both the $x$ and $y$ directions, as shown in
Fig.2a.

\begin{figure} \includegraphics[scale=0.18]{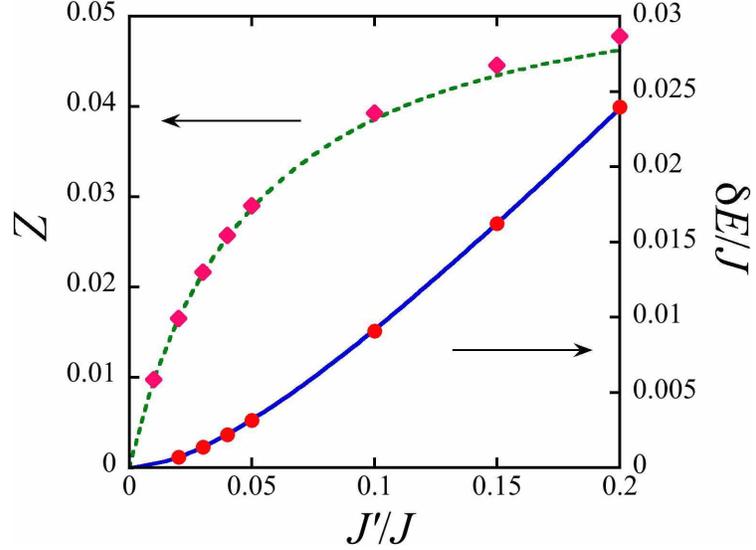} \caption{(Left
$y$-axis) Transition rate to the bound state ($Z$). Pink diamonds are
obtained by the present approach, and the green dotted line is the
analytical result in Eq.\eqref{eq:19}. The small deviation in the
large $J'/J$ regime is due to the correction to the ground state which
is not included in Eq.\eqref{eq:19}. (Right $y$-axis) Gap between the
bound state and the lower edge of continuum ($\delta E$). Red circles
are obtained by the present approach, and the blue solid line denotes
the RPA result. The data shown here are calculated at ${\bf
k}=(\pi/4,\pi)$ on a spatially anisotropic triangular lattice with
$J'_1=J'_2(\equiv J')$ and $J'_3=0$. } \label{fig:RPA} \end{figure}

To compare with the bound state data from the effective Schr\"odinger
equation \eqref{eq:2}, we calculate $S_{1D}$ and $\chi'$ numerically.
Results for the binding energy $\delta E = \omega_{2,l} - \omega_t$
obtained in these two calculations are found to essentially coincide
with each other as shown in Fig. \ref{fig:RPA}. We also observe that
in the regions where the width of the 2-spinon continuum becomes
comparable to $\delta E$, the scaling changes to $\delta E \sim
J'({\bf k})$.  The change from quadratic to linear scaling can be
understood simply from \eqref{eq:chiB}, and simply corresponds to the
situations where the argument of the cotangent is large (quadratic) or
$O(1)$ (linear).  We also analyze spectral weight (residue) $Z$ of the
triplon, see \eqref{eq:19}. We find that for small $J'$ it scales as
$\sqrt{\delta E} \propto J'$. A comparison between the numerical and
analytical results in a wider range of inter-chain exchange values is
shown in Fig.\ref{fig:RPA}.

It should be mentioned that the multiplicative logarithmic factor in
$\chi'$, eq.\eqref{eq:chiB}, does lead to a weak ``spiral''
instability at zero frequency -- i.e. in this approach one finds a
bound state with negative energy at some ordering momentum (measured
from $\pi$)\cite{RPA_Cs2CuCl4}.  This instability, however, is {\sl
  very} weak. The corresponding ordering momentum is extremely small,
$k_{x,0} \sim 10^{-10}$ \cite{RPA_Cs2CuCl4}, translating into
similarly tiny ``instability energy'' $\sim J k_{x,0}$.  Moreover,
this classical instability is overshadowed by a stronger quantum (of
collinear type) ones, of the order $(J')^4/J^3$ \cite{triangDimer}.
Since our approximation is concerned with the features of the dynamical
structure factor at energies of order $J'$ and higher, we are allowed to
disregard these weak instabilities.

{\sl Anti-bound states} are analyzed similarly. In this case,
$\omega_t({\bf k}) = \omega_{2,u}(k_x) + \delta_t({\bf k})$ is above
the 2-spinon continuum where $\chi' < 0$, see \eqref{eq:chi}.  Since
$\chi'(\omega)$ continuously decreases in magnitude to zero but
retains the same (negative) sign as $\omega$ in increased from
$\omega_{2u}$ to $\infty$, the condition for the existence of an
anti-bound state is simply $1 + J'({\bf k}) \chi'(k_x,\omega_{2,u})
<0$.  The anti-bound state therefore merges into the two-spinon
continuum when $1 + J'_{\rm crit}({\bf k}) \chi'(k_x,\omega_{2,u}) = 0$.  
As $J'({\bf k})$ is increased above this threshold, one can show
that, because $S({\bf k},\omega) \sim {\rm Const}\sqrt{\omega_{2u}-\omega}$ 
for $\omega\lesssim \omega_{2u}$ \cite{chainMqwcalc}, the anti-bound
state energy scales as $\delta_t({\bf k}) \propto (J'({\bf k}) -
J'_{\rm crit}({\bf k}))^2$ while its spectral weight scales as
$\sqrt{\delta_t}$, similarly to the bound state situation discussed
above. Unlike the bound state, the anti-bound one is not a completely sharp
excitation.  This is because it takes place in the region of 4-spinon
excitations, which extend from $\omega_{2,l}(k_x)$ up to
$\omega_{4,u}(k_x) = \pi J \sqrt{2(1 + |\cos [k_x/2]|)}$, \cite{caux}.
Hence, $\chi''\neq 0$ and the triplon lineshape is in fact Lorentian,
see \eqref{eq:Scontfinal}.  However, the 4-spinon spectral weight is
very small in the region between $\omega_{2,u}$ and $\omega_{4,u}$
boundaries \cite{caux}, and we find that anti-bound states remain well
defined, with the height-to-width ratio well above $1$.

Being a collective excitation above the two-particle continuum, the
anti-bound triplon here is very similar to the familiar zero-sound
mode in a Fermi-liquid, e.g. ${}^3$He. The analogy is made much more precise by
focusing on the region near $\Gamma$ point in the Brillouin zone,
where the 2-spinon continuum collapses onto a line: $\omega_{2,u} -
\omega_{2,l} \sim k_x^3$ as $k_x \to 0$. In this region
$\chi'(k_x,\omega) = |k_x|/(2(v_s |k_x| - \omega))$, where $v_s = \pi
J/2$ is spinon velocity.  The dispersion is found immediately (see
Fig.2b):
\begin{equation}
\label{eq:abs}
\omega_t({\bf k}) = v_s |k_x| \Big(1 + \frac{J'({\bf k})}{2v_s}\Big) ~ 
{\rm for}~ J'({\bf k}) > 0 .
\end{equation}
We see that the anti-bound triplon is just an ``acoustic plasmon'' of  
the spinon gas
with short-range interactions.

As anti-bound states away from the $\Gamma$ point require strong $J'$
for their existence, we would like to suggest that somewhat more
two-dimensional spatially anisotropic triangular antiferromagnet
Cs$_2$CuBr$_4$ \cite {CsCuBr} seems to be a promising candidate for
the corresponding inelastic neutron scattering study.

%\bibliography{apssamp}

\newpage

\end{document}